\documentclass[twocolumn,amsmath,amssymb,aps,prc,nofootinbib]{revtex4-2}
\usepackage{graphicx,here}
\usepackage{dcolumn,bm,hyperref,footnote}
\usepackage{float}
\usepackage[dvipsnames,svgnames,x11names]{xcolor}
\usepackage[authormarkup=none,defaultcolor=BrickRed]{changes}
\usepackage{multirow}

\begin{document}

\title{Classical and Bayesian error analysis 
of the relativistic mean-field model for doubly magic nuclei}

\author{M. Imbri{\v s}ak}
\email{mimbrisa@phy.hr}
\affiliation{Department of Physics, Faculty of Science, 
University of Zagreb, HR-10000 Zagreb, Croatia}

\author{K. Nomura}
\email{nomura@nucl.sci.hokudai.ac.jp}
\affiliation{Department of Physics, 
Hokkaido University, Sapporo 060-0810, Japan}
\affiliation{Nuclear Reaction Data Center, 
Hokkaido University, Sapporo 060-0810, Japan}

\date{\today}

\begin{abstract}
The information-geometric statistical analysis 
on the stability of model reductions, 
reported previously 
[Imbri{\v s}ak and Nomura, 
Phys. Rev. C {\textbf{107}}, 034304 (2023)] 
with a focus on 
the manifold boundary approximation method 
in the application to the nuclear density-dependent 
point-coupling model of infinite nuclear matter, 
is extended to the 
numerically more challenging case of finite nuclei. 
A simple procedure is presented for 
determining the binding energies 
of doubly magic nuclei within the 
relativistic mean-field framework using 
the Woods-Saxon potential. 
The proposed procedure, employing the 
Fisher information matrix 
combined with algorithmic differentiation, 
is shown to provide reliable estimates 
of parameter uncertainties of the nuclear 
energy density functional 
for finite nuclei, while reducing  
the time-consuming sampling 
of the parameter space, which would 
be required in the numerically more 
involved Bayesian statistical techniques. 
\end{abstract}

\maketitle

\section{Introduction\label{sec:intro}}

The nuclear energy density functionals (EDFs) 
are a widely used framework for describing 
nuclear structure phenomena. 
Many such EDFs are based on the relativistic mean-field (RMF) 
Lagrangian in the finite-range meson-exchange model 
\citep{Bender2003Self-consistentStructure}. 
The density-dependent meson-nucleon couplings 
have been successfully applied in this framework 
to describe asymmetric nuclear matter 
\citep{VRETENAR2005RelativisticStructure}. 
Alternatively, since the exchange of heavy mesons cannot be resolved at low energies, the self-consistent RMF framework can be formulated in terms of 
point-coupling nucleon interactions. 
This approach yields comparable results 
to the meson-exchange coupling approach for 
finite nuclei \citep{Nikolaus1992NuclearModel,Burvenich2002NuclearModel}.  
For example, the successful phenomenological finite-range 
interaction, denoted as the density-dependent meson-exchange (DD-ME2), 
is mapped to the point-coupling 
framework by relating the strength parameter of
the isoscalar-scalar derivative term to different values of the mass
of the phenomenological $\sigma$ meson in the DD-ME2 model \citep{Niksic2008Finite-Interactions}. 
The resulting ``best-fit model'', 
such as the density-dependent point-coupling 
(DD-PC1) functional (see e.g., \citep{Niksic.2015}), 
requires the fine-tuning of the density dependence 
of the isoscalar-scalar and isovector-vector 
interaction terms to nuclear matter 
and ground-state properties of finite nuclei.  

The issue of uncertainty quantification and error propagation in nuclear EDFs has recently attracted attention, focusing on the study of error estimates by statistical analysis \citep{Piekarewicz2014InformationPhysics, Schunck2015UncertaintyTheory},
assessment of systematic errors \citep{Dobaczewski2014ErrorGuide, Schunck2014ErrorTheory}, and correlation analysis \citep{Reinhard2010TheSkin, Schunck2014ErrorTheory}.  
However, statistical analysis is more challenging 
for the point-coupling models since they are found 
to exhibit an exponential range of sensitivity 
to parameter variations \citep{Niksic.2015}. 
This behavior is found to be a feature 
of imprecise models, that is, models 
that depend only on a few
rigidly constrained combinations of 
the parameters \citep{Waterfall.2006}.

Recent advancements 
\citep{Transtrum.2011,Machta2013ParameterModels, Transtrum.2015,Transtrum.2016} 
in the understanding of the behavior 
of model uncertainties have yielded new approaches, 
such as the manifold boundary approximation method (MBAM) 
\citep{Transtrum.2014}. 
MBAM is a systematic procedure for reducing 
model uncertainties by constructing progressively 
more precise lower-dimensional models from 
an initial imprecise higher-dimensional model. 
This construction is based on the concepts from 
information geometry - an interdisciplinary field 
that introduces differential geometry concepts 
to statistical problems 
\citep{Amari1982,Amari2016InformationApplications}.

MBAM has already been used to systematically 
construct effective nuclear density functionals 
of successively lower dimensions and smaller 
impact of uncertainty. This method was illustrated 
on the DD-PC1 functional evaluated for pseudodata 
for infinite symmetric nuclear matter 
in Ref.~\citep[][]{Niksic.2016}. 
In Ref.~\cite{Niksic.2017}, this analysis was extended to calculate the derivatives of observables with respect to model parameters, and it has become possible to apply the MBAM to realistic models constrained not only by the pseudo-data related to the nuclear matter equation of state but also by observables measured in finite nuclei. 
In our recent paper \citep{imbrisak2023} 
we investigated the overall stability of the MBAM procedure applied in the reduction of nuclear structure models using methods of information geometry and Monte Carlo 
simulations. 
In the illustrative application 
to the DD-PC1 model of the nuclear EDF, we found that the main conclusions 
obtained by using the MBAM method are stable under 
the variation of the parameters within the $1\sigma$ 
confidence interval of the 
best-fitting model.

In contrast to the simple case of infinite nuclear matter, where one would have to solve only a simple iterative procedure to obtain the Dirac mass and binding energy, finite nuclei require a careful description of the nuclear many-body problem. 
Broadly speaking, statistical analysis can be 
performed either in the Bayesian framework, i.e., 
by employing elaborate Monte Carlo simulations, 
or in the ``classical'' framework, 
found by computing the Fisher information matrix (FIM) and its inverse (the covariance matrix) 
from the chosen statistical model 
(see, e.g., \citep[]{Schunck2014ErrorTheory}). 
The latter approach should be, in principle, less time-consuming than running an extensive Monte Carlo simulation. 
However, when computing the FIM, one has to constrain the first derivatives of the chosen statistical model, either numerically or analytically. 
Attempting a simple extension of existing implementations 
of RMF FORTRAN codes 
\cite{Gambhir1990,Gambhir1993,Ring1997,Niksic2014} 
would introduce uncertainties due to employing 
numerical differentiation. 
We, therefore, intend to implement 
a simple proof-of-concept 
version of a finite nucleus code in PYTHON, 
in which well-tested libraries for 
algorithmic differentiation (AD) exist.

The analysis presented below is based on a procedure for determining the RMF binding energies, starting from a simple and widespread \cite{Ring1997,Niksic2014} assumption of a Woods-Saxon potential, often used to compute the starting point for density-dependent potentials. This paper compares numerically estimating parameter errors using a chosen Bayesian statistical technique - the Markov chain Monte Carlo (MCMC) to the faster method of directly determining the covariance matrix without sampling but using the AD-determined FIM.
 
The paper is organized as follows. 
In Sec.~\ref{sec:rmf}, we give an overview of 
the RMF procedure implemented in the present analysis, and 
in Sec.~\ref{sec:data} we describe the inputs 
used for our PYTHON routines. 
In Sec.~\ref{sec:results} we present the results 
of our statistical analysis.
Finally, conclusion is given 
in Sec.~\ref{sec:conclusion}.

\section{Numerical implementation of the RMF procedure\label{sec:rmf}}

In this section, we briefly overview the overall 
relativistic Lagrangian (Sec.~\ref{sec:lag}) and 
the chosen pairing model (Sec.~\ref{sec:pairing}). 
We describe 
the matrix elements for the Dirac equation 
for the proton and neutron single-particle energies 
in the spherical system (Sec.~\ref{sec:spherical}). 
and the functional form of the Woods-Saxon 
potential that is implemented in our 
PYTHON codes (Sec.~\ref{sec:WSPot}).

\subsection{Relativistic Lagrangian\label{sec:lag}}

The relativistic Lagrangian governing point-coupling 
models is based on a set of bilinear currents 
\begin{equation}
\bar{\psi}\mathcal{O}_{\tau}\Gamma\psi, \mathcal{O}_{\tau}\in\{1,\tau_i\}, \Gamma\in\{1,\gamma_\mu,\gamma_5,\gamma_5\gamma_\mu,\sigma_{\mu\nu}\} \; ,
\end{equation}
where $\psi$ is the Dirac spinor, 
used to describe nucleons, $\tau_i$'s 
are the Pauli matrices for 
isospin, and $\Gamma$ represents the Dirac matrices. 
The resulting Lagrangian may be divided into 
the free-particle $\mathcal{L}_{free}$, 
bilinear current $\mathcal{L}_{4f}$, 
bilinear current derivative $\mathcal{L}_{der}$, 
and the electromagnetic $\mathcal{L}_{em}$, 
components \cite{Finelli2004RelativisticSymmetry}: 
\begin{equation}
\mathcal{L}=\mathcal{L}_{free}+\mathcal{L}_{4f}+\mathcal{L}_{der}+\mathcal{L}_{em}.\label{eq:Luk}
\end{equation}
The interacting parts of the Lagrangian are 
composed of four types of fermion interactions: 
the isoscalar-scalar $(\bar{\psi}\psi)^2$, 
isovector-vector 
$(\bar{\psi}\gamma_\mu\psi)(\bar{\psi}\gamma^\mu\psi)$,
isovector-scalar  
$(\bar{\psi}\vec{\tau}\psi)\cdot(\bar{\psi}\vec{\tau}\psi)$, 
and the isovector-vector type  
$(\bar{\psi}\vec{\tau}\gamma_\mu\psi)\cdot(\bar{\psi}\vec{\tau}\gamma^\mu\psi)$.

In the point-coupling models, the interacting terms are added to the Lagrangian by multiplying the bilinear currents by their respective couplings (denoted by $\delta_S$,  $\alpha_S$, $\alpha_V$, $\alpha_{TS}$ and $\alpha_{TV}$) that are dependent on the baryon density, $\hat\rho$, defined as
\begin{equation}
\hat{\rho} u^\mu=\bar{\psi}\gamma^\mu\psi \; ,
\end{equation}
where $u^\mu$ is the four-velocity 
$u^\mu=(1-v^2)^{-1/2}(1,\vec{v})$. 
The considered class of point-coupling models employs only second-order terms, disregarding, e.g., six-fermion and eight-fermion vertices but, instead, promotes the coupling constants to functions of nucleon density \cite{Finelli2004RelativisticSymmetry}. 
These models are built on the same building blocks 
as in the meson-exchange models, 
wherein the single-particle properties are tied 
to the three meson fields: 
the isoscalar-scalar $\sigma$ meson, 
the isoscalar-vector $\omega$ meson, and 
the isovector-vector $\rho$ meson, 
without the isovector-scalar term 
\cite{Niksic2008Finite-Interactions}.

\subsection{Pairing}\label{sec:pairing}

Pairing is a crucial nuclear correlation in open-shell 
nuclei and is, therefore, necessary to describe 
nuclei that are not doubly magic \cite{Ring1996}. 
Although it is not necessary to include pairing for 
the set of doubly magic nuclei, we do not restrict 
our codes in that manner. This is to ensure 
that the analysis presented below can be easily 
extended to future work dealing with open-shell 
nuclei where pairing correlations play a role. 
In the constant gap approximation \cite{Vautherin1973}, 
each single-particle state is occupied according 
to the occupation probability, $v_i^2$, 
calculated by using the BCS formula
\begin{equation}
v_i^2=\frac{1}{2}
\left[1-\frac{\epsilon_i-\lambda}{\sqrt{(\epsilon_i-\lambda)^2+\Delta^2}}
\right] \; ,
\label{eq:BCS_v2}
\end{equation}
where $\lambda$ is the chemical potential and 
$\Delta$ is the gap parameter. 
The chemical potential is determined separately 
for protons and neutrons by finding a solution 
to the equations for the chemical potentials 
for protons and neutrons, 
\begin{align}
\sum\limits_i v_{i,p}^2(\lambda_p)&=Z\\
\sum\limits_i v_{i,n}^2(\lambda_n)&=N \; ,
\label{eq:BCS_v2_norms}
\end{align}
so that the total numbers of neutrons and protons are conserved.
The pairing energy can then be computed 
from a simple expression 
\begin{equation}
    E_{pair}=-G\sum\limits_i (v_i u_i)^2 \; ,
\end{equation}
where $u_i$ is the unoccupation amplitude satisfying 
$u_i^2=1-v_i^2$, and $G$ is a constant 
determined from the self-consistency condition
\begin{align}
    \Delta = G \sum\limits_i u_i v_i.
\end{align}
Since the sum necessary for computing the 
pairing energy diverges, one often introduces 
cutoff energy \cite{Ring1996,Gambhir1990}.

\subsection{The spherical system}\label{sec:spherical}

The procedure is based on solving the Dirac equation 
for the single-particle energies for protons and neutrons 
in the spherical system. 
First, the single-particle wavefunction is decomposed 
into the isospin wavefunction $\chi_{t_i}(t)$, 
the spin wavefunction $\chi_{1/2}(s)$, 
the angular momentum wavefunction, $Y_l(\theta,\phi)$, 
and two spinor radial components, $f(r)$ and $g(r)$. 
Due to symmetry considerations, the solutions are 
separable in terms of the total angular momentum 
$j$, and parity $\pi$, 
yielding the following relations: 
\begin{align}
    l(j,\pi)&=j+\pi/2 \; , \\
    \widetilde l(j,\pi)&=j-\pi/2 \; , \\
    \kappa(j,\pi) &= \pi (j+1/2) \; .
\end{align} 
The maximal radial quantum number needs to be truncated in practical calculations to obtain finite matrices.
The maximum radial quantum number for the expansion of radial functions $f$ and $g$ ($n_{max}$ and $\widetilde n_{max}$, respectively) are determined as functions of the final major shell quantum number $N_F$. The value of the maximal radial quantum number of the function $g$ is greater than the maximal value for $f$ to avoid spurious solutions. These states of a high radial quantum number close to the Fermi surface arise from the lack of coupling for the $f_{n_{max}}$ state to the $g$ states through the $\mathbf{\sigma}\cdot\nabla$ term when a truncation of the quantum number is applied 
\cite{Gambhir1990,Gambhir1993,Ring1997}, i.e., 
\begin{align}
    n_{max}&=\frac{N_F-l(j,\pi)}{2}\\
    \widetilde n_{max}&= N_F+1 \; .
\end{align}
In this separation, a joint spin and 
angular momentum quantum numbers, $|ljm\rangle$, 
are represented with the two-dimensional spinor
\begin{equation}
\Phi_{ljm}(\theta,\phi,s) = 
\left[\chi_{1/2}(s)\otimes Y_l(\theta,\phi)
\right]_{jm} \; .
\end{equation}
The full wavefunction can then be written as
\begin{equation}
    \psi(r,\theta,\phi, s,t)=
\begin{pmatrix}
                f(r)\Phi_{ljm}(\theta,\phi,s)\\
                i g(r)\Phi_{\widetilde l jm}(\theta,\phi,s)
    \end{pmatrix} \; .
\end{equation}

After separating the isospin, spin, 
and angular momentum components, 
one can use the simplified Hamiltonian 
for a single $(j,\pi)$ block for protons 
and neutrons, whose solution depends 
only on the radial coordinate 
\begin{equation}
    \psi_{j\pi}(r)=\begin{pmatrix}
                f_{j\pi}(r)\\
                i g_{j\pi}(r)
    \end{pmatrix} \; .
\end{equation}
Both $f$ and $g$ functions are expanded 
using the relativistic quantum 
harmonic oscillator basis
\begin{equation}
    R_{n,l}=N_{n,l}L_n^{l+1/2}(\xi^2)\xi^le^{-\xi^2/2} \; ,
\label{eq:Rnl_functions}
\end{equation}
where the radial coordinate has been rescaled 
to a dimensionless quantity $\xi$ using the 
scaling parameter $b_0=\sqrt{1.011 A^{1/3}}$. 
The expansion includes a finite range of 
radial quantum numbers that are different 
for $f$ and $g$ functions
\begin{equation}
    \begin{pmatrix}
f\\ g\end{pmatrix}=
\begin{pmatrix}
\sum\limits_{n}^{n_{max}} f_n R_{n,l}\\
\sum\limits_{\widetilde n}^{\widetilde{n}_{max}} g_{\widetilde{n}} R_{\widetilde{n},\widetilde{l}}
\end{pmatrix} \; .
\end{equation}
The limits $n_{max}$ and $\widetilde n_{max}$ 
are dependent on the total quantum number 
$N_F$ and angular momentum. 

For each $(j,\pi)$ block, 
the Dirac equation is solved using the effective mass $M$ and potential $V$. 
The aforementioned ansatz, 
$\psi=(f(r),ig(r))$, 
yields the following matrix equation:
\begin{equation}
    \begin{pmatrix}
V+M-m & \hbar c\left(\partial_r-\frac{\kappa-1}{r}\right)\\
-\hbar c\left(\partial_r+\frac{\kappa+1}{r}\right) & V-M-m
\end{pmatrix}
\begin{pmatrix}
f \\ g
\end{pmatrix}_{j\pi} = \epsilon
\begin{pmatrix}
f \\ g
\end{pmatrix}_{j\pi}.
\end{equation}
Using the relativistic harmonic oscillator basis introduced 
in Eq.~(\ref{eq:Rnl_functions}), 
this matrix equation can be structured as
\begin{equation}
    \begin{pmatrix} A & B^T\\
                B & -C \end{pmatrix} \begin{pmatrix}f_1\\\vdots\\ g_{\widetilde n_{max}}\end{pmatrix}=\epsilon \begin{pmatrix}f_1\\\vdots\\ g_{\widetilde n_{max}}\end{pmatrix},
\end{equation}
using three matrices $A_{nn'}$, $B_{\widetilde n,n'}$ and $C_{\widetilde n, \widetilde n'}$:
\begin{align}A_{n,n'} &= \int\limits_0^\infty r^2 dr R_{n,l} R_{n',l}(V+M-m) \\
B_{\widetilde n,n'} &= \hbar c\int\limits_0^\infty r^2 dr R_{\widetilde n,\widetilde l} \left(-\partial_r-\frac{\kappa+1}{r}\right) R_{n',l} \\
C_{\widetilde n,\widetilde n'} &= \int\limits_0^\infty r^2 dr R_{\widetilde n,\widetilde l} R_{\widetilde n',\widetilde l}(M+m-V) .
\end{align}
Once the wave functions are known, 
the pairing is introduced as an additional 
weight to the density of each eigenstate, 
$v_i^2$, as outlined in Sec.~\ref{sec:lag} 
using Eq.~(\ref{eq:BCS_v2_norms}).

\subsection{The Woods-Saxon potential}\label{sec:WSPot}

We shall apply the finite nucleus procedure 
to the simple case of the Woods-Saxon potential. 
The Woods-Saxon potential is also the first 
step for more complex density-dependent potentials. 
The shape of the Woods-Saxon potential is known, and this potential does not depend on the nucleon densities. 
Therefore, unlike density-dependent potentials, 
the procedure need not be run iteratively, 
thus reducing computational complexity 
for various numerical tests. 

The shape of the potential has been 
adapted from \cite{Koepf1991}, 
the authors of which developed a relativistic equivalent of the simple Woods-Saxon potential. 
In their model, a set of 12 parameters 
was used to constrain the shape of the Woods-Saxon 
potential by describing both the potential 
and the effective mass. 
Their model accomplishes this by introducing 
four different potentials - the normal 
($U_p$ and $U_n$) and spin-orbit potentials 
($W_p$ and $W_n$) for protons and neutrons. 
These potentials were tied to the 
vector, $V$, and scalar, $S$, potentials 
in the Dirac equation by considering their 
nonrelativistic limit as
\begin{align}
    U &= V+S \\
    W &= V-S \; .
\end{align}
The strengths of all four potentials are regulated 
by the overall potential strength $V_0$ and modulating 
factors for different numbers of protons and neutrons, 
$\kappa$, and for the strength of the spin-orbit 
contribution, $\lambda_p$ and $\lambda_n$. 
The shape of the potentials is regulated by four diffusivities, $a_p$, $a_n$, $a^{ls}_{p}$ and $a^{ls}_{n}$, and four radii $R_0^n$, $R_0^p$, $R_{0,ls}^n$, and $R_{0,ls}^p$. \footnote{The notation of \cite{Koepf1991} has been simplified and the signature of the spin-orbit potentials has been absorbed into $\lambda_n$ and $\lambda_p$ for convenience.}
The resulting potentials are given as follows:
\begin{align}
U_p(r) 
&=
\frac{V_0\left(1+\kappa \frac{N-Z}{A}\right)}
{1+e^{\frac{r-R_0^p A^{1/3}}{a_p}}}+U_C(r)
\\
U_n (r)
& =\frac{ V_0 \left(1-\kappa \frac{N-Z}{A}\right)}{1+e^{\frac{r-R_0^n A^{1/3}}{a_n}}}
\\
W_p(r) &=\frac{V_0\lambda_{p}\left(1+\kappa \frac{N-Z}{A}\right)}{1+e^{\frac{r-R_{0,ls}^p A^{1/3}}{a^{ls}_p}}}+W_C(r)\\
    W_n(r) & =  \frac{V_0\lambda_{n}\left(1-\kappa \frac{N-Z}{A}\right)}{1+e^{\frac{r-R_{0,ls}^n A^{1/3}}{a^{ls}_n}}}.
\end{align}
An additional component describing the repulsive 
Coulomb potential $U_C$, is added to the potential 
of protons using the homogeneously charged sphere potential
\begin{align}
    U_C(r) &= \begin{cases}Z e^2\left(\frac{3}{R_0^p A^{1/3}}-\frac{r^2}{(R_0^p)^3 A}\right), & r\leq R_0^p A^{1/3}\\
    \frac{Z e^2}{r}, & r>R_0^p A^{1/3}\end{cases}\\
    W_C(r) &= \begin{cases}Z e^2\left(\frac{3}{R_{0}^p A^{1/3}}-\frac{r^2}{(R_{0}^p)^3 A}\right), & r\leq R_{0}^p A^{1/3}\\
    \frac{Z e^2}{r}, & r>R_{0}^p A^{1/3}\end{cases}.
\end{align}

\subsection{Fisher information matrix\label{sec:info-geom}}

We want to compute error estimates for the problem 
of fitting a model $f^a(\mathbf{p})$ to measurements $y^a$, 
assuming measurement errors $\sigma^a$. 
Here, we use indices from the beginning of the 
Latin alphabet for $N_m$ measurements, 
and the Greek letters for $N_p$ model parameters 
[labeled as $\mathbf{p}=(p^1,\cdots,p^{N_p})$].  
In the standard maximum likelihood method, the best-fitting value of $p^\mu$ is found by minimizing the $\chi^2$ value, 
\begin{equation}
 \chi^2(\mathbf{p})=\sum\limits_{a=1}^{N_m} \left(\frac{y^a-f^a(\mathbf{p})}{\sigma^a}\right)^2 \; .
\end{equation}
A useful derived quantity is the reduced 
$\chi^2$ value $\chi^2_{red}=\chi^2/(N_m-N_p)$, 
which should be close to $1$ for models that 
are neither overfitted nor underfitted. 

We find parameter uncertainties 
using the Cramer-Rao bound on the covariance matrix $\sigma$, 
which is based on the inverse of the FIM, denoted by $g_{\mu\nu}$
\citep{Amari2016InformationApplications}:
\begin{equation}
g_{\mu\nu}(\mathbf{p})= \sum\limits_a \frac{ \partial_\mu f^a\partial_\nu f^a}{\left(\sigma^a\right)^2} \; .
\end{equation}
We compute model derivatives using algorithmic
differentiation implemented in the AUTOGRAD 
package. Using AD procedures, we eliminate 
numerical errors related to using numerical differentiation approximations.

\section{Input selection}\label{sec:data}

\begin{figure}[ht]
\begin{center}
\includegraphics[width=\linewidth]{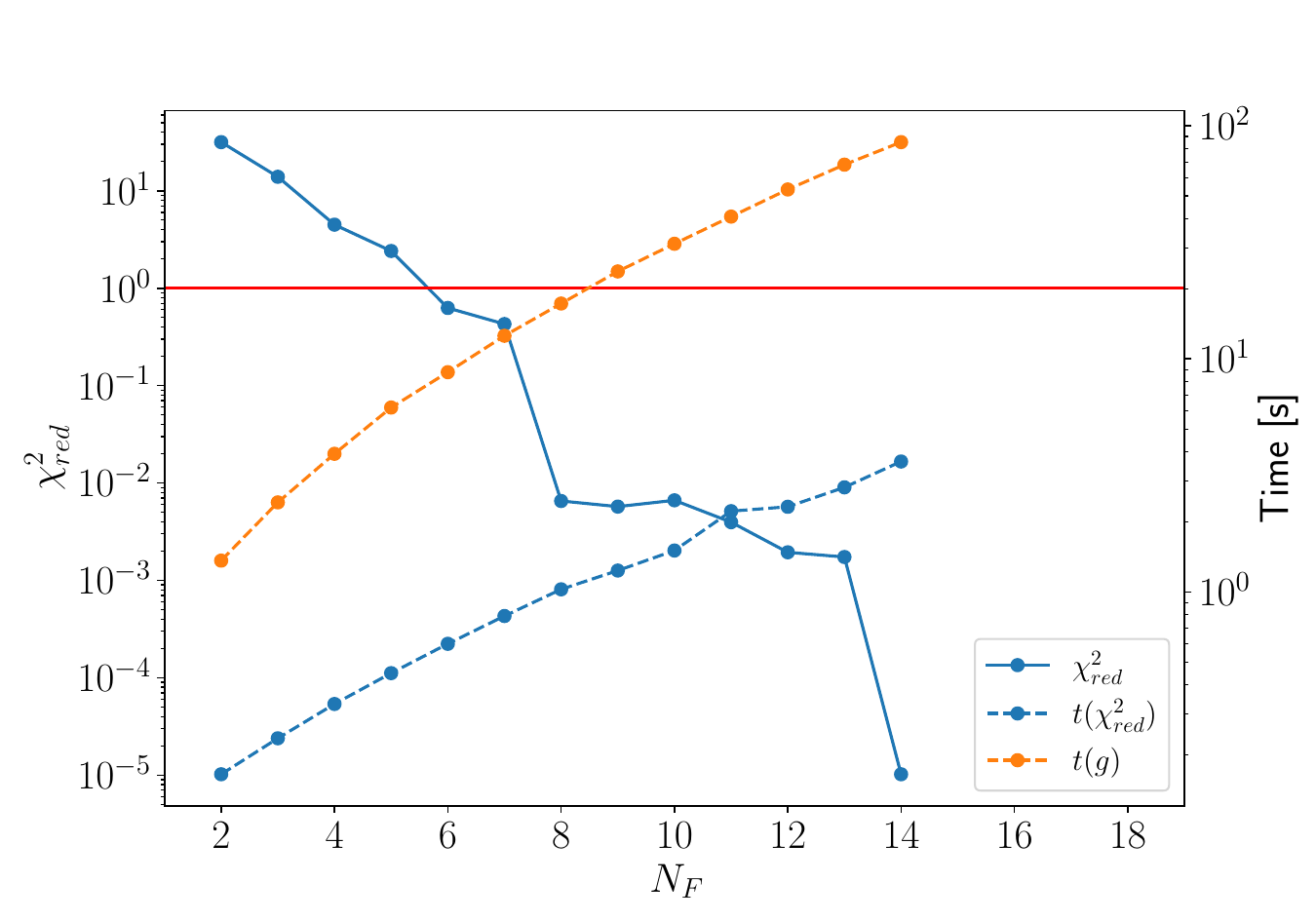}
\caption{Reduced $\chi^2_{red}$ value of the 
finite-nucleus model as a function of $N_F$ 
for the Woods-Saxon potential. 
The dashed lines represent the execution time 
of the $\chi^2_{red}$ function and the computation 
time of the Woods-Saxon FIM. }
\label{fig:WS_optimal}
\end{center}
\end{figure}

\begin{table}[h]
\caption{\label{tab:finite_dataset}
The data set consisting of the 
charge radii, $r_{ch}$, and single neutron, 
$\epsilon_n$, and proton, $\epsilon_p$, energies 
for occupied states. 
The single-particle energies are computed using 
the Woods-Saxon potential as determined 
in Ref.~\cite{Koepf1991}.}
\begin{ruledtabular}
\begin{tabular}{cccccccc}
Nucleus & $r_{ch}$ (fm) \\
\colrule
$^4$He    &1.65 & & & &       &         &       \\
$^{16}$O  &2.41 & & & &       &         &       \\
$^{40}$Ca &3.29 & & & &       &         &       \\
\colrule
 & \multicolumn{6}{c}{$\epsilon_n$ (MeV)} \\
\cline{2-7}
 & $1s_{1/2}$ & $1p_{3/2}$ & $1p_{1/2}$ & $2s_{1/2}$ 
 & $1d_{5/2}$ & $1d_{3/2}$ \\
\colrule
  $^4$He    &$-$25.30 & &        &        &       &         &       \\
   $^{16}$O & $-$43.20 &$-$24.68 &$-$19.04&&&&\\
$^{40}$Ca   &$-$53.34 &$-$39.40 &$-$35.40 &$-$24.95 &$-$18.51 &$-$17.42\\
\colrule
 & \multicolumn{6}{c}{$\epsilon_p$ (MeV)} \\
\cline{2-7}
 & $1s_{1/2}$ & $1p_{3/2}$ & $1p_{1/2}$ & $2s_{1/2}$ 
 & $1d_{5/2}$ & $1d_{3/2}$ \\
\colrule
 $^4$He &$-$24.95 &        &        &        &        &       \\
   $^{16}$O&$-$40.08 &$-$22.39 &$-$18.36 &&&\\
$^{40}$Ca &$-$45.80 &$-$33.08 &$-$30.32 &$-$19.53 &$-$14.96 &$-$13.23\\
\end{tabular}
\end{ruledtabular}
\end{table}

\begin{figure*}[ht]
\begin{center}
\includegraphics[width=\linewidth]{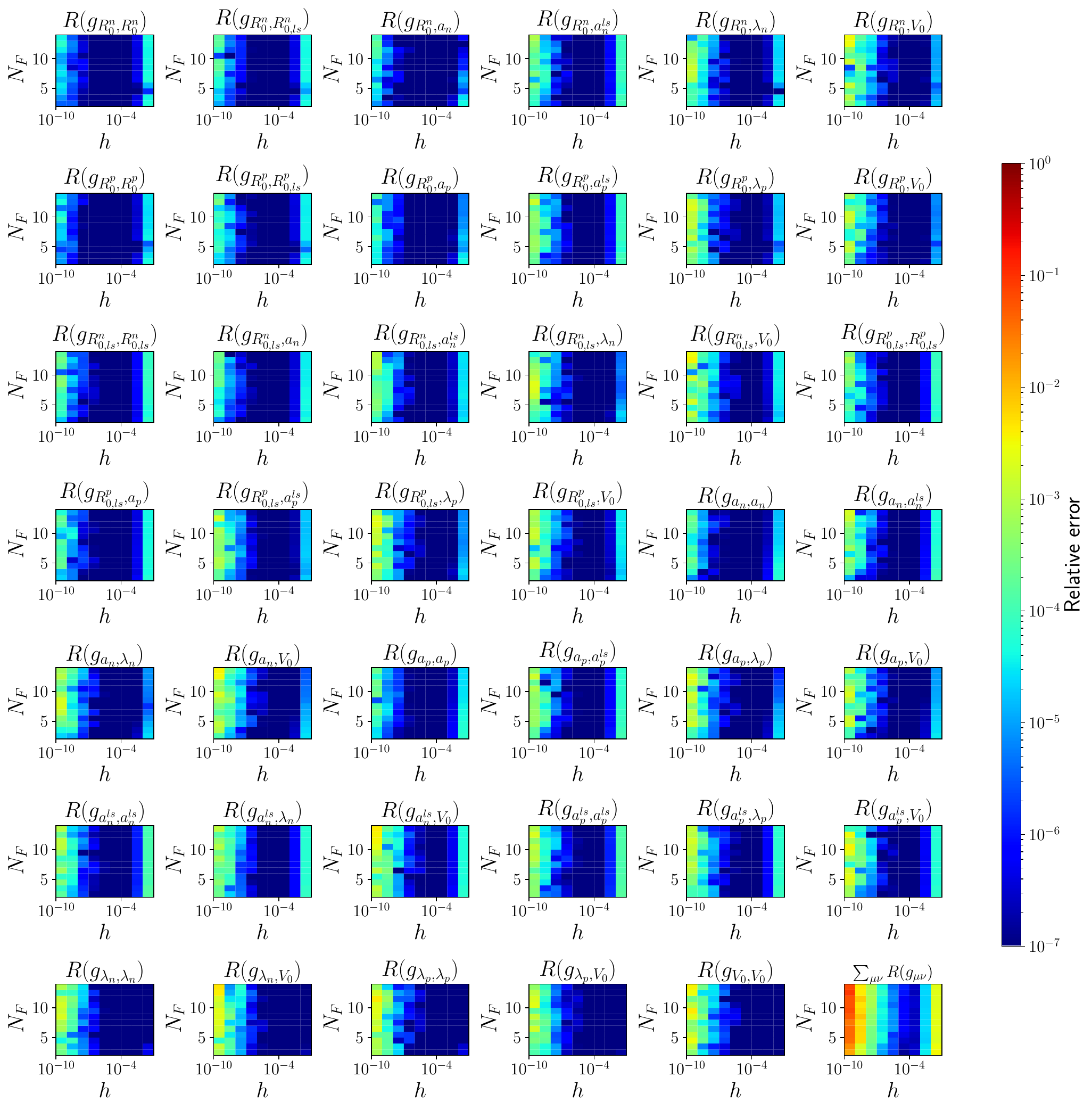}
\caption{Relative error of the different 
FIM components, color-coded as a function of $N_F$ 
and numerical derivative step $h$. 
The relative error compares the AD-derived FIM estimate, 
$g_{\mu\nu}^{(A)}$, 
to the numerical estimate $g_{\mu\nu}^{(N)}$. 
On the bottom-right panel, the sum of all relative 
errors, $\sum_{\mu\nu}R(g_{\mu\nu})$, is plotted.}
\label{fig:WS_optimal_g}
\end{center}
\end{figure*}

We analyze the statistical properties of the 
RMF procedure on charge-radius, $r_{ch}$, and single-particle 
energy data. To this end, we choose a set of 
doubly magic nuclei: $^4$He,  $^{16}$O, and $^{40}$Ca. 
Since the parameter space consists 
of 12 parameters and only three nuclei, the chosen data set consists of their charge-radii and the single-particle energies of protons and neutrons for occupied states, computed using the values of Ref.~\cite{Koepf1991}. 
For statistical analyses, these parameter values were taken as the best-fitting values for the Woods-Saxon potential.

Using charge radii and the energies of the occupied single-particle states results in 23 data points, ensuring enough degrees of freedom for a twelve-parameter model. 
A further simplification comes from the fact that the proton and neutron numbers are the same in all three nuclei, excluding the parameter $\kappa$ and thereby reducing the parameter space to 11 dimensions. 
We compute the charge-radius, $r_{ch}$, 
from the root-mean-square radius, 
$\langle r^2\rangle$, (as in, e.g., \cite{Niksic2014}) 
using the proton density distribution,  as $r_{ch}=\sqrt{\langle r^2\rangle+0.64}$.
A homoscedastic error of $0.1\,\mathrm{fm}$ and $0.1\,\mathrm{MeV}$ has been chosen arbitrarily since the data set consists 
of the model evaluation and not of spectral measurements. 
The difference in charge radius and single-particle energy values is smaller than the chosen error.

The corresponding reduced $\chi^2_{red}$ value of the finite-nucleus model as a function of $N_F$ for the Woods-Saxon potential is shown in Fig. \ref{fig:WS_optimal}. The choice of a different error would only shift the $\chi^2_{red}$ 
curve upwards or downwards. 
The figure also shows the execution time 
as a function of the maximal total quantum number $N_F$, 
displayed as a dashed line. 
The simple relation $\chi^2_{red} \approx 1$ should hold to minimize the impact of overfitting and underfitting. 
The model accomplishes this near $N_F \approx 5$. 
Since the execution time of the $\chi^2_{red}$ function rises progressively with a larger $N_F$, the value of the $N_F$ parameter is set to 5 for statistical analyses.  
The execution time for the FIM matrix for this model shows similar behavior. 
The chosen data set is shown in Table~\ref{tab:finite_dataset} 
and is computed using a $N_F=15$, 
which is set outside the examined $N_F$ range 
in Fig. \ref{fig:WS_optimal} in order to avoid 
the artificial $\chi^2=0$ data point. 
The value of $N_F$ is chosen to be large enough 
so that the values of all computed parameters 
differ by less than $10\%$ of the adopted value 
for the homoscedastic error between neighboring values of $N_F$.

\begin{figure*}[ht]
\begin{center}
 \includegraphics[width=\linewidth]{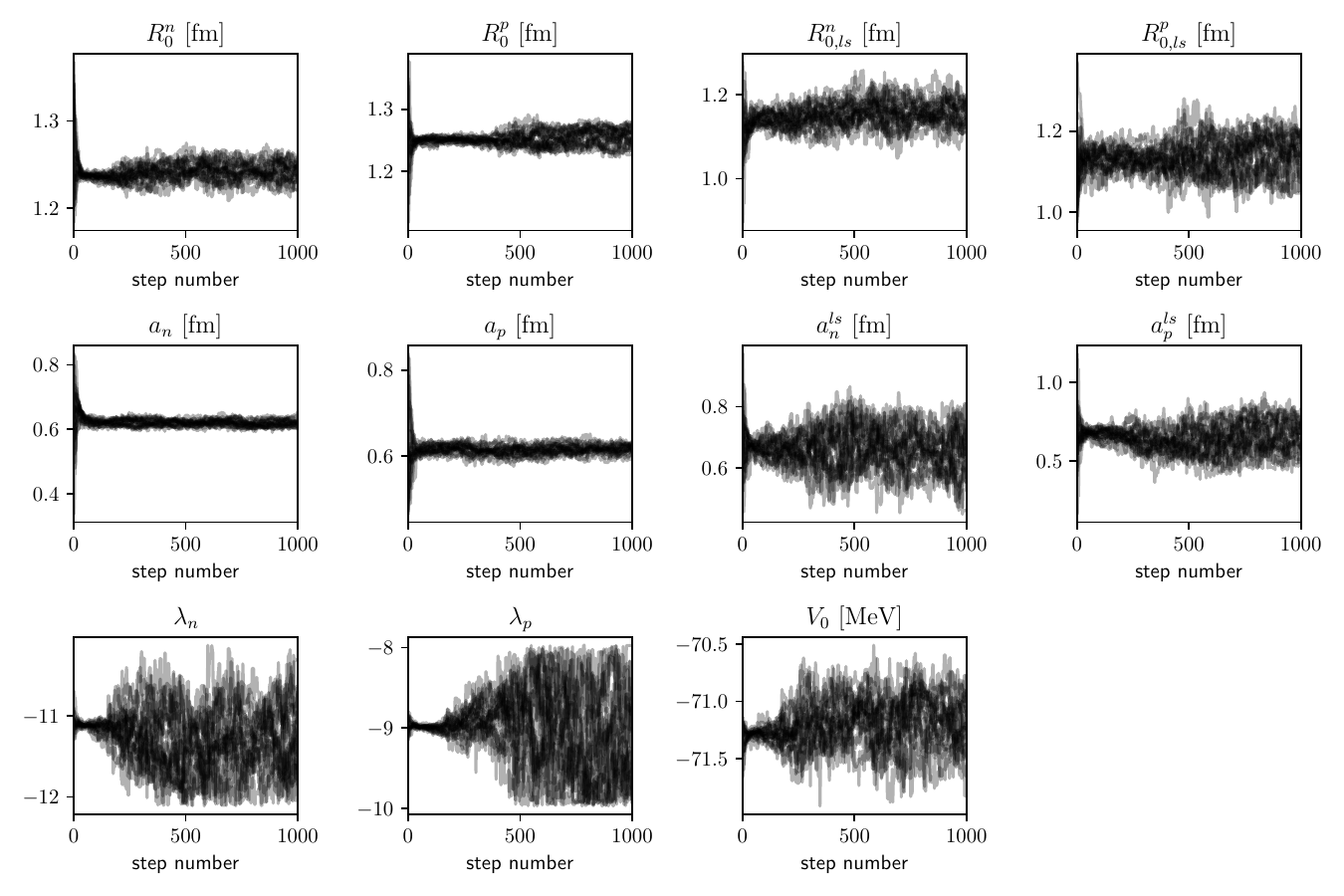}
\caption{Values of the individual Markov chains of 
the MCMC sampling as a function of the MCMC step. }
\label{fig:MCMC_finite_nuclei_chains}
\end{center}
\end{figure*}

\begin{figure*}[ht]
\begin{center}
 \includegraphics[width=\linewidth]{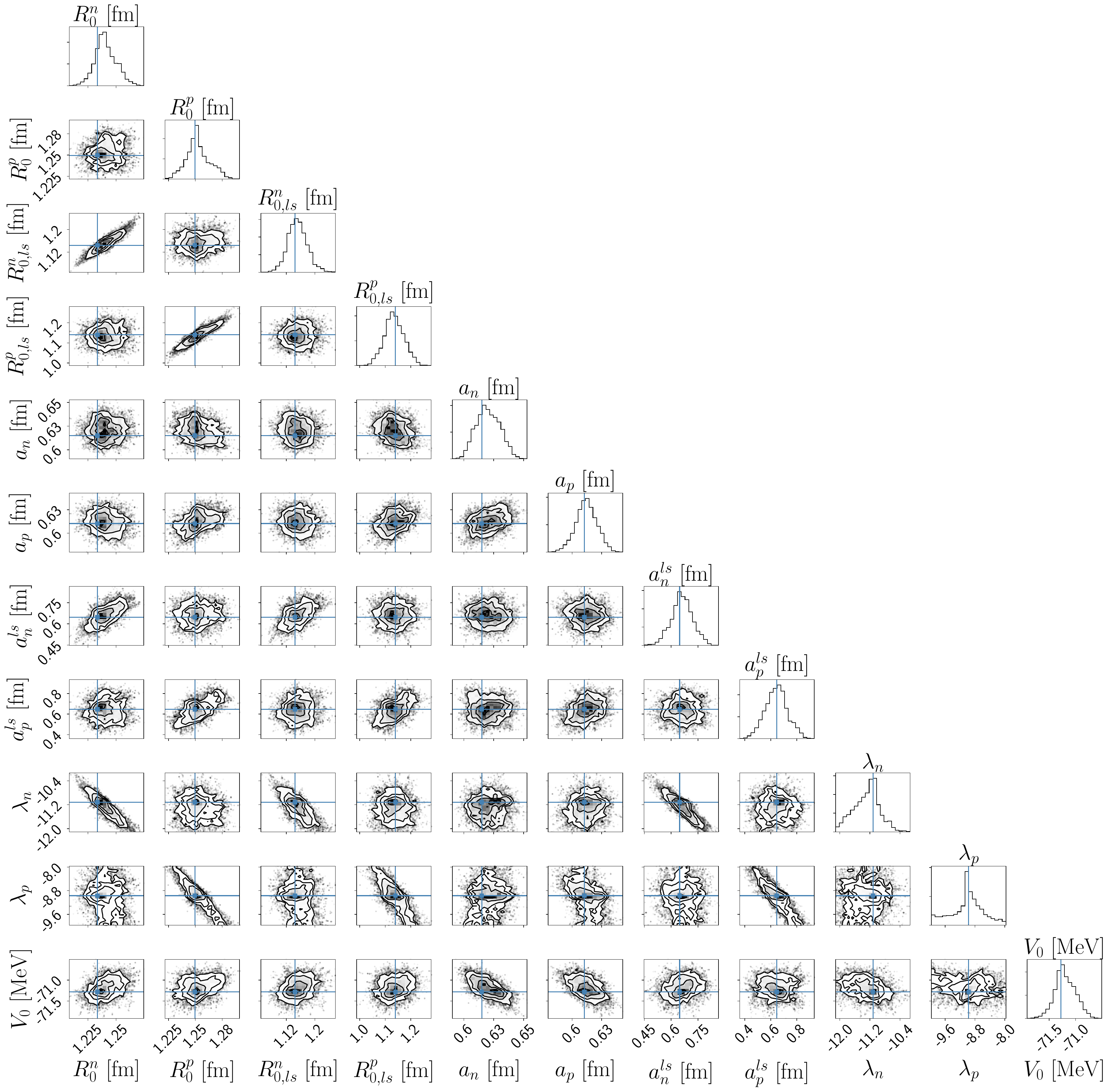}
\caption{MCMC-derived sampling of the 
Woods-Saxon potential shown as two-dimensional 
sections of the parameter space. }
\label{fig:MCMC_finite_nuclei}
\end{center}
\end{figure*}

\begin{table*}[ht!]
\caption{\label{tab:MCMC_WS_results}
Results of the Woods-Saxon potential fitting 
using the FIM-based techniques and the MCMC 
method for the charge radius 
and single-particle energy data set 
given in Table~\ref{tab:finite_dataset}. 
In the last two columns shown are $Z$ scores 
defined in Eq.~(\ref{eq:zscores}).}
\begin{ruledtabular}
\begin{tabular}{lccccccc}
 Parameter & Ref.~\cite{Koepf1991} & $\sigma_\textnormal{MCMC}$ 
& $\sigma_\textnormal{FIM}$ & MCMC interval & Confidence 
& $|Z(\sigma_\textnormal{MCMC})|$ & $|Z(\sigma_\textnormal{FIM})|$ \\
\colrule 
$R_0^n$ (fm) & 1.2334 &   0.0095 &   0.0112 & $1.24 \pm 0.01$ & [ 1.23,    1.25 ] &     0.79 &     0.66\\
$R_0^p$ (fm) & 1.2496 &   0.0108 &   0.0168 & $ 1.25 \pm  0.01$ & [ 1.24,    1.26 ] & 0.28 &     0.18\\
$R_{0,ls}^n$ (fm) & 1.1443 &   0.0273 & 0.0320 & $ 1.15 \pm  0.03$ & [ 1.13,    1.18 ] &     0.27 &     0.23\\
$R_{0,ls}^p$ (fm) & 1.1401 &   0.0389 &   0.0563 & $ 1.14 \pm  0.04$ & [ 1.10,    1.18 ] &     0.11 &     0.07\\
$a_{n}$ (fm) & 0.6150 &   0.0097 &   0.0098 & $ 0.62 \pm  0.01$ & [ 0.61, 0.63 ] &     0.56 &     0.55\\
$a_{p}$ (fm) & 0.6124 &   0.0107 &   0.0108 & $ 0.61 \pm  0.01$ & [ 0.60, 0.63 ] &     0.20 &     0.20\\
$a^{ls}_{n}$ (fm) & 0.6476 &   0.0601 &   0.0746 & $ 0.66 \pm  0.06$ & [ 0.60,    0.72 ] &     0.23 &     0.18\\
$a^{ls}_{p}$ (fm) & 0.6469 &   0.0848 &   0.1271 & $ 0.64 \pm  0.08$ & [ 0.56,    0.72 ] &     0.05 &     0.03\\
$\lambda_n$ & $-$11.1175 &   0.3391 &   0.4167 & $-11.3 \pm   0.3$ & [ $-$11.65,  -10.98 ] & 0.49 &     0.39\\
$\lambda_p$ & $-$8.9698 &   0.4287 &   0.7025 & $-9.0 \pm   0.4$ & [ $-$9.47, $-$8.61 ] &     0.07 &     0.04\\
$V_0$ (MeV) & $-$71.2800 &   0.1941 &   0.2228 & $-71.2 \pm 0.2$ & [ $-$71.37, $-$70.99 ] & 0.45 &     0.39\\
$\kappa$ & 0.4616 &  &    &  &  &  &  \\
\end{tabular}
\end{ruledtabular}
\end{table*}

\section{Results}\label{sec:results}

We apply the finite nucleus procedure to compute 
parameter uncertainties for the Woods-Saxon potential. 
We estimate errors of the model parameters 
obtained by computing the diagonal elements 
of the FIM, $\sigma_\textnormal{FIM}$, as presented 
in Table~\ref{tab:MCMC_WS_results}. 
Below we compare the values of the FIM components 
computed using AD and those computed using numerical 
differentiation. We also compare the FIM-derived 
error estimates to those of the MCMC technique.

The numerical 
differentiation is compared to the one that employs 
a symmetric differentiation step $h$. 
Figure \ref{fig:WS_optimal_g} shows 
the relative error, $R$, for the different 
components of the FIM, which is computed as
\begin{equation}
    R(g_{\mu\nu})=\left|\frac{g_{\mu\nu}^{(A)}-g_{\mu\nu}^{(N)}}{g_{\mu\nu}^{(A)}}\right| \; .
\end{equation}
Here, $g_{\mu\nu}^{(A)}$ is our AD-derived FIM 
estimate of the $\mu\nu$ matrix component of the FIM, 
and $g_{\mu\nu}^{(N)}$ is the numerical estimate 
computed with a differentiation step $h$. 
In Fig.~\ref{fig:WS_optimal_g} we show these relative 
errors computed for different values of $h$ and $N_F$. 
For very small values of $h<10^{-7}$ the 
numerical errors due to floating point precision 
accumulate, while for $h>10^{-2}$ the finite 
difference approximation tends to break down. 
This behavior is observed for all $N_F$, and the relative error values do not depend strongly on $N_F$. 
To assess the overall worst-case error scenario, 
we compute the sum of all relative errors, 
$\sum_{\mu\nu} R(g_{\mu\nu})$. 
This quantity is shown in the bottom right panel 
of Fig.~\ref{fig:WS_optimal_g}, and 
suggests that the optimal $h$ is consistently 
$h \approx 10^{-4}$ for the entire range of $N_F$. 
We conclude that the AD implementation provides 
accurate estimates of the FIM and that any 
discrepancy to the numerical derivative can 
be attributed to the inherent issues 
of numerical derivatives.

We use the MCMC technique to sample the 
$\chi^2$ posterior distribution, 
as implemented in the package EMCEE 
\cite{2013PASP..125..306F}. 
We use samples of 24 Markov chains of length 1000. 
The number of initialized chains is chosen to fulfill the MCMC requirement that the number of Markov chain 
walkers be greater than the number of dimensions 
of the parameter space. 
In Fig.~\ref{fig:MCMC_finite_nuclei_chains} 
we plot the values of the MCMC samples of the parameter 
space as a function of the step in the Markov chain 
in which they are produced. 
We see that the values stabilize after $\approx 50$ initial steps, 
indicating the expected \emph{burn-in} 
phase for the MCMC method \cite{2013PASP..125..306F}. 
The sampled data points corresponding to the initial 50 steps are excluded from further analysis.

In Fig.~\ref{fig:MCMC_finite_nuclei} we show 
both the two-dimensional and 
one-dimensional marginal distributions of the 
MCMC samples in the parameter space. 
The blue lines show the value expected from the literature, 
which is well aligned with the distribution of 
the MCMC samples in all panels in 
Fig.~\ref{fig:MCMC_finite_nuclei}. 
The error estimates computed with MCMC sampling 
are listed alongside the FIM-based technique in 
Table \ref{tab:MCMC_WS_results}. 
The medians and the $1\sigma$ confidence interval 
derived with the MCMC sampling are well aligned with 
the estimates in Ref.~\cite{Koepf1991}. 
To assess the degree of 
statistical differences between the parameters 
estimated in Ref.~\cite{Koepf1991}, $p^\mu_0$, 
and the MCMC-based best-fitting parameter 
values of our data set, 
$p^\mu$, we list in the last two columns 
of Table~\ref{tab:MCMC_WS_results} the $Z$ scores, 
defined by 
\begin{eqnarray}
\label{eq:zscores}
 Z^\mu(\sigma)=\frac{p^\mu_{0}-p^\mu}{\sigma} \; .
\end{eqnarray}
We find that the differences are generally not 
statistically significant (i.e., they are less than $1\sigma$), 
regardless of whether 
$\sigma_\textnormal{FIM}$ or $\sigma_\textnormal{MCMC}$ 
is considered. 
The FIM method is based on the assumption 
of the Gaussianity of errors, while the MCMC algorithm 
is not constrained in that way. 
Using longer MCMC chains would remove 
the remaining numerical differences between 
$\sigma_\textnormal{FIM}$ and $\sigma_\textnormal{MCMC}$ 
in Table~\ref{tab:MCMC_WS_results}. 
Choosing to compute error estimates via MCMC sampling 
over FIM would only be useful if there is a need 
to analyze the impact of non-Gaussian errors. 

In contrast to simply considering the diagonal elements of the FIM inverse, by reliably computing the FIM with the aid of AD, one can perform error analysis 
without invoking the time-consuming sampling of the parameter space. 
The resulting estimates, $\sigma_\textnormal{FIM}$, are 
in agreement with the MCMC estimates, $\sigma_\textnormal{MCMC}$, 
as shown in Table~\ref{tab:MCMC_WS_results}.

\section{Conclusion\label{sec:conclusion}}

Uncertainties related to parameter estimation 
for nuclear EDFs have recently become 
a topic under rigorous investigations. 
By extending our previous study of \citep{imbrisak2023}, 
in which methods of information geometry 
was applied to EDFs in the case of infinite 
nuclear matter, in this work we have presented a 
statistical analysis of a simple procedure for determining 
the RMF binding energies for a set of doubly-magic nuclei 
with the Woods-Saxon potential. 
We have compared error estimates between the faster 
procedure that employs the FIM and the numerically more 
challenging Bayesian MCMC method.  
Even in the complex case of finite nuclei, 
EDF parameter uncertainties can be reliably estimated 
by using the FIM combined with algorithmic differentiation. 
The proposed approach to error analysis has 
the advantage of avoiding the time-consuming sampling 
of the parameter space, which would otherwise 
be required in Bayesian statistical techniques.

In our next step, by using the optimized 
Woods-Saxon potential resulting from the 
present analysis, 
we will apply nuclear structure codes 
to give error estimates for the point-coupling 
models, such as the universal 
EDF DD-PC1, for 
a realistic study of finite nuclei. 
Work along this line is in progress, and 
will be reported elsewhere.

\acknowledgments
The work of M.I. is financed within 
the Tenure Track Pilot Programme of 
the Croatian Science Foundation and 
the \'Ecole Polytechnique F\'ed\'erale de Lausanne, 
and the project TTP-2018-07-3554 
Exotic Nuclear Structure and Dynamics, 
with funds from the Croatian-Swiss Research Programme.

\bibliography{references.bib}

\end{document}